\documentclass[prl,amsmath,amssymb,floatfix,twocolumn]{revtex4}

\usepackage{graphicx}% Include figure files
\usepackage{dcolumn}% Align table columns on decimal point
\usepackage{bm}% bold math
\usepackage{psfrag}
\usepackage{subfigure}

\newcommand{\be}{\begin{eqnarray}}
\newcommand{\ee}{\end{eqnarray}}
\newcommand{\bn}{\begin{eqnarray*}}
\newcommand{\en}{\end{eqnarray*}}

\newcommand{\nn}{\nonumber \\}
\newcommand{\nl}{\\}

\renewcommand{\vec}[1]{\mbox{\boldmath$#1$}}
\renewcommand{\d}{\mbox{\rm d}}
\newcommand{\gslash}[1]{\mbox{\slash{\hspace{-2mm}}$#1$}}

\renewcommand{\th}{\ensuremath{\theta}}

\newcommand{\vph}{\ensuremath{\varphi}}

\newcommand{\sg}{\ensuremath{\sigma}}

\newcommand{\Cx}{\ensuremath{\hat{x}}}
\newcommand{\Cy}{\ensuremath{\hat{y}}}

\newcommand{\sgvec}{\ensuremath{\vec{\sg}}}

\newcommand{\lt}{\ensuremath{\left}}
\newcommand{\rt}{\ensuremath{\right}}

\newcommand{\PhiG}{\ensuremath{\Phi_{\rm G}}}

\renewcommand{\k}{\ensuremath{k_0}}

\renewcommand{\d}{\mbox{\rm d}}

\begin{document}

%\preprint{APS/123-QED}

%\title{Manuscript Title:\\with Forced Linebreak}% Force line breaks with \\

%\author{Ann  Author}
%\affiliation{%

%\affiliation{


%\preprint{APS/123-QED}

\title{Reply to Comment on ``Can gravity distinguish between \\ Dirac and Majorana neutrinos?''}%

%\author{Dinesh Singh}
%\email{singhd@uregina.ca}
%\author{Nader Mobed}
%\email{nader.mobed@uregina.ca}
%\author{Giorgio Papini}
%\email{papini@uregina.ca}
%\altaffiliation[Also at ]{Prairie Particle Physics Institute, Regina, Saskatchewan, S4S 0A2, Canada;
%International Institute for Advanced Scientific Studies, 89019 Vietri sul Mare (SA), Italy.}
%\affiliation{Department of Physics, University of Regina \\ Regina, Saskatchewan, S4S 0A2, Canada}

\author{Dinesh Singh$^{a}$}
\altaffiliation[Electronic address:  ]{singhd@uregina.ca}
\author{Nader Mobed$^{a}$}
\altaffiliation[Electronic address:  ]{nader.mobed@uregina.ca}
\author{Giorgio Papini$^{a,b,c}$}
\altaffiliation[Electronic address:  ]{papini@uregina.ca}
\address{$^a$Department of Physics, University of Regina, Regina, Saskatchewan, S4S 0A2, Canada}
\address{$^b$Prairie Particle Physics Institute, Regina, Saskatchewan, S4S 0A2, Canada}
\address{$^c$International Institute for Advanced Scientific Studies, 89019 Vietri sul Mare (SA), Italy}

\date{\today}

%\begin{abstract}

%We show that spin-gravity interaction can distinguish between Dirac and Majorana
%neutrino wave packets propagating in a Lense-Thirring background.
%Using time-independent perturbation theory and gravitational phase to generate a perturbation Hamiltonian with spin-gravity coupling,
%we show that the associated matrix element for the Majorana neutrino differs significantly from its Dirac counterpart.
%This difference can be demonstrated through significant gravitational corrections to the neutrino oscillation length for
%a two-flavour system, as shown explicitly for SN1987A.
%%(Draft only.)

%\end{abstract}

%\pacs{04.90.+e, 14.60.Pq, 04.80.+z, 97.60.Bw}
%%\pacs{Valid PACS appear here}

\maketitle

%\author{D. Singh}
%\email{singhd@uregina.ca}
%\affiliation{%
%Department of Physics, University of Regina \\
%Regina, Saskatchewan, S4S 0A2, Canada
%}%
%\author{N. Mobed}
%\email{nader.mobed@uregina.ca}
%\affiliation{%
%Department of Physics, University of Regina \\
%Regina, Saskatchewan, S4S 0A2, Canada
%}%
%\author{G. Papini}
%\email{papini@uregina.ca}
%\affiliation{%
%Department of Physics, University of Regina \\
%Regina, Saskatchewan, S4S 0A2, Canada
%}%
%\affiliation{
%International Institute for Advanced Scientific Studies, 89019 Vietri sul Mare (SA), Italy.}
% %Lines break automatically or can be forced with \\


%\begin{abstract}

%\end{abstract}

%\maketitle

%\section{\label{sec:level1}First-level heading:\protect\\ The line
%break was forced \lowercase{via} \textbackslash\textbackslash}

%\subsection{\label{sec:level2}Second-level heading: Formatting}
%\subsubsection{\label{sec:level3}Third-level heading: References and Footnotes}

This comment is in response to Nieves and Pal \cite{Nieves} who dispute our claim \cite{Singh-PRL}
that a classical gravitational field may possibly distinguish between Dirac and Majorana neutrinos,
described in terms of Gaussian wave packets propagating in a Lense-Thirring background,
where the distinction is manifested in spin-gravity corrections to the neutrino oscillation length.
They contend that our model for Majorana neutrinos is incorrect and that any
conclusions relating to this neutrino type are not reliable.
%, though they raise no objections
%with our construction for Dirac neutrinos.
Furthermore, they suggest that any distinction between the two neutrino types will be suppressed
by factors of $m/E$ where $m$ is the neutrino mass and $E$ is its mean energy of propagation,
and claim that this distinction is unobservable when $m/E \ll 1$.
%which should vanish in the zero-mass limit.
%While we wish to first thank Nieves and Pal for their interest in our Letter and appreciate their feedback,
%we have many reasons to
We beg to disagree.

%The first category concerns the technical points they raise.
As noted above, our model for the Dirac and Majorana neutrinos is motivated by a wave packet approach
in {\em quantum mechanics}, as opposed to a plane-wave expansion in {\em quantum field theory}.
This is an essential detail which Nieves and Pal %unfortunately
fail to acknowledge.
In addition, the gravitational field \cite{Singh-PRL} is incorporated in terms of a gravitational phase $\PhiG$,
giving rise to an interaction Hamiltonian $H_{\PhiG}$ with spin-dependent features, to be evaluated in terms of
time-independent perturbation theory.
These two details are important for framing the context underpinning our reply.
Regarding the technical concerns, we agree that the Majorana condition they note
in their eq. (1) is certainly true for a {\em fermion field operator}.
However, we again emphasize that our perspective is {\em quantum mechanical}, so our treatment
of the Majorana condition must be described in terms of {\em wave functions}.
Adopting their notation, our approach is to identify \cite{Aitchison,Fukugita} $u_R$ with $(u_L)^c = i \, \sg^2 \, u_L^*$
and also separately identify $u_L$ with $(u_R)^c = -i \, \sg^2 \, u_R^*$ to
obtain \cite{error}
\be
W_1 & = & u_L + (u_L)^c, \qquad W_2 \ = \ u_R - (u_R)^c.
\label{W1,W2}
\ee

Unlike what Nieves and Pal claim, it indeed follows \cite{Aitchison} that $W_{1,2}$ is a solution of the free particle equation
$\gslash{k} \, W_{1,2} = \pm \, m \, W_{1,2}$ \cite{error} for wave functions.
%
%Having stated that, we do acknowledge that our construction (\ref{W1,W2}) was imprecisely implemented.
%Following our notation \cite{Singh-PRL}, we present a more complete Majorana wave packet model, in the form
However, a more precise implementation of (\ref{W1,W2}) with our notation leads to a Majorana wave packet model with the form
\be
| \psi_{1(2)} \rangle^{\rm Maj.} & = &
{1  \over (2\pi)^{3/2}} \int \d^3 k \, \xi(\vec{k}) \, |W_{1(2)}(\vec{k})\rangle^{\rm Maj.},
\label{psi=}
\ee
where
\be
|W_1(\vec{k})\rangle^{\rm Maj.} & = & e^{-i k \cdot x} | \nu_L \rangle + e^{i k \cdot x} | \nu_L^c \rangle,
\label{W1-mod}
\nl
|W_2(\vec{k})\rangle^{\rm Maj.} & = & e^{-i k \cdot x} | \nu_R \rangle - e^{i k \cdot x} | \nu_R^c \rangle,
\label{W2-mod}
\ee
and $\xi(\vec{k})$ is the Gaussian function \cite{Singh-PRL}.
Clearly, it follows from (\ref{psi=}) that $| \psi_{1(2)}^c \rangle^{\rm Maj.} = \pm \, | \psi_{1(2)} \rangle^{\rm Maj.}$.
This leads to a modification of our eq. (13) for the Majorana matrix element \cite{Singh-PRL}, which is
\be
\lefteqn{\langle \psi_{1(2)}(\vec{r}) |H_{\Phi_{\rm G}}| \psi_{1(2)}(\vec{r}) \rangle^{\rm Maj.} \ = \
\langle \psi(\vec{r}) |H_{\Phi_{\rm G}}| \psi(\vec{r}) \rangle^{\rm Dirac}
}
\nn
&&{} \pm (\hbar \, \k) \sin \th \, \sin \vph
%\nn
%&&{} \times
\lt[ {M \over r}
\langle \pm |\sgvec|\mp \rangle^{\Cy} \lt[C_{0 \Cy} + C_{1 \Cy} \, \bar{m} + C_{2 \Cy} \, \bar{m}^2\rt] \rt.
\nn
&&{} + \lt. {M \Omega R^2\over r^2} \langle \pm |\sgvec|\mp \rangle^{\Cx} \lt[D_{0 \Cx} + D_{1 \Cx} \, \bar{m} + D_{2 \Cx} \, \bar{m}^2\rt] \rt].
\label{amplitude-Maj}
\ee

Our plots \cite{Singh-PRL}, as applied to the SN~1987A data, are {\em completely unaffected} by the adjustments
because the corrections only alter the contributions coupled to $M \Omega R^2 / r^2$,
which are all exponentially damped compared to the $M/r$ contributions.
As for the $m/E$ suppression issue raised by Nieves and Pal, this is of no relevance because
all such terms are {\em automatically excluded} within the construction of $|W_{1,2}(\vec{k})\rangle^{\rm Maj.}$,
so everything presented in \cite{Singh-PRL} is of leading order.
As shown in (\ref{amplitude-Maj}), and also present in our eq. (13) of \cite{Singh-PRL},
Dirac and Majorana neutrinos behave differently under the gravitational perturbation $\PhiG$ by way of the
spin-flip terms $\langle \pm |\sgvec|\mp \rangle^{\Cx,\Cy}$.
A spin-flip changes a Majorana neutrino into an antineutrino and behaves like a charge conjugation operation.

%The second category concerns the philosophical reasons for not using standard quantum field theory
%to investigate this problem.
%This is because, despite much research on quantum field theory in curved space-time \cite{Birrell},
%there are many conceptual difficulties in precisely understanding how classical gravity interacts
%with quantum matter.
%The wave packet approach allows for the quantum particle to sense changes in space-time curvature
%while still retaining a point-like quality to avoid problems associated with back-reaction and
%other highly non-trivial challenges.
%This model is extremely useful for obtaining actual predictions that may become sensitive for
%verification by future neutrino experiments, thereby giving us a greater insight into the nature
%of gravity at the quantum mechanical level.
%We respectfully disagree with Nieves and Pal that a gravitational interaction should be treated
%with the same considerations as that of, for example, the electroweak interaction, since
%it is not clear whether Einstein's theory is well-behaved at the Compton wavelength scale, and whether
%this behaviour has any impact upon determining a future quantum gravity theory that can
%be experimentally tested.

%%%%%%%%%%%%%%%%%%%%%%%%%%%%%%%%%%%%%%%%%%%%%%%%%%%%%%%%%%%%%%%%%%%%%%%%%%%%%%%%%%%%%%%%%%%%%%%

\end{document}